\documentclass{aa}
\usepackage{txfonts}
\usepackage{graphicx}
\usepackage{natbib}
\bibpunct{(}{)}{;}{a}{}{,}

\def\hdsyv{HD\,27536}
\def\hdeen{HD\,216803}
\def\kms{km\,s$^{-1}$}

\begin{document}

\title{Binarity, activity and metallicity among late-type stars\thanks{Based on observations collected at the 
    La Silla Observatory, ESO (Chile), with the HARPS
    spectrograph at the ESO 3.6m telescope.}}
\subtitle{I. Methodology and application to HD 27536 and HD 216803}

\author{T. H. Dall\inst{1}
    \and
    H. Bruntt\inst{2}
    \and
    K. G. Strassmeier\inst{3}
}

\offprints{T. H. Dall, \email{tdall@eso.org}}

\institute{European Southern Observatory, Casilla 19001, Santiago 19, Chile
    \and
    Niels Bohr Institute, Juliane Maries Vej 30, 2100 Copenhagen \O, Denmark
    \and
    Astrophysical Institute Potsdam (AIP), An der Sternwarte 16, D-14482 Potsdam, Germany
}

\date{Received / Accepted }

\abstract{We present the first in a series of papers that attempt to
investigate the relation between binarity, magnetic activity, and chemical
surface abundances of cool stars. In the current paper, we lay out
and test two abundance analysis methods and apply them to two
well-known, active, single stars, \object{HD 27536} (G8IV-III) and
\object{HD 216803} (K5V), presenting 
photospheric fundamental parameters and abundances
of Li, Al, Ca, Si, Sc, Ti, V, Cr, Fe, Co and Ni. The abundances from the two methods agree
within the errors for all elements except calcium in \hdeen ,
which means that either method yields the same fundamental model
parameters and the same abundances. Activity is described by the
radiative loss in the Ca\,{\sc ii} H\&K lines with respect to the
bolometric luminosity, through the activity index $R_\mathrm{HK}$. Binarity is established by very precise
radial velocity (RV) measurements using HARPS spectra. The spectral line bisectors
are examined for correlations between RV and bisector shape
to distinguish between the effects of stellar activity and unseen companions.
We show that HD\,27536 exhibit RV variations mimicking the effect of a
low-mass ($m \sim 4$~M$_J$) companion in a relatively close ($a \sim 1$~AU) orbit.
The variation is strongly correlated with the activity, and consistent with the known photometric period
$P = 306.9$~d,
demonstrating a remarkable coherence between $R_\mathrm{HK}$ and the bisector shape, i.e. between the photosphere and the chromosphere.
We discuss the complications involved in distinguishing between companion and
activity induced RV variations.

 }

\maketitle

\section{Introduction}

There is currently still no single theory that adequately
describes magnetic activity on cool main-sequence stars with outer
convective envelopes. Neither is there a solid description of its
impact on stellar evolution. While the current scheme of dynamo
models successfully account for the existence of magnetic fields
for stars at the age of the Sun, more quantitative explanations
are lacking \citep[see][for a recent review]{strassmeier2003}.

It has been established empirically \citep{noyes+1984} that the
activity level, and hence the dynamo efficiency, is a monotonic
function of the rotation period for stars with periods larger then
a few days. It scales approximately with the Rossby number
$\mathcal{R}_0 = P \tau_c^{-1}$, where $\tau_c$ is the convective
turnover time and $P$ the rotational period. Hence, the activity
is indirectly a function of the spectral type. However, the
boundary layer dynamo model is unable to account for the activity
in the overactive M dwarfs, and also the reason for and the exact
scaling with $\mathcal{R}_0$ is not fully understood
\citep[e.g.][]{schrijver1996,schrijver+zwaan2000}.

Although rotation is thought to be the necessary condition for
magnetic activity, it is not clear to what extent binarity
influences the generation and morphology of magnetic fields and
the corresponding chromospheric and coronal emission
\citep[e.g.][]{bopp+1977,edwards1983,shkolnik+2003}. The
differential gravitational pull from a companion may cause a
longitude- and latitude-dependent relationship between rotation
rate and activity level, 
and may also contribute to an inhomogeneous chemical abundance by 
introducing tidal waves which affect the chemical stratification. No theoretical studies exist in the literature, but
if such relationships exist, the models of the evolution of close
binaries would then need to be reconsidered.

Another burning question is the link between chemical surface
abundances and binarity. The recent observation that (solar-like)
stars with planets are on average more metal rich than similar
stars without planets \citep[e.g.][]{santos+2001,santos+2004,santos+2005,fischer+2003}
is at least suggestive that this maybe also the case for binaries.
Of particular interest is the fact that the Li abundance seem to differ between single stars and stars which host
giant planets \citep{israelian+2004}, and even between otherwise identical twin-star binaries \citep{martin+2002,dall+2005b}, possibly
related to pre-MS accretion of planetesimals or engulfment of planets.

Furthermore, as instrumentation improves  \citep[i.e.~HARPS --
see][]{harps2003}, more and more single stars turn out to have
variable radial velocity (RV), due to the presence of low-mass
companions, either low-mass stars, brown dwarfs or giant planets.
It is thus becoming necessary and indeed possible to deal properly
with the effects of binarity, and to attempt to separate the
classical dynamo process of internal differential rotation from
the  external contribution from close companions.  In this
discussion it is necessary to extend the concept of binarity to
include all possible companions, including brown dwarfs and
planets.

We are currently conducting a search for true single stars among
the  known active stars, in order to study the activity-rotation
relation in a sample that is not ``polluted'' by any type of
binaries. Furthermore, we may hope to isolate the effects that
binarity may have on stellar activity. Our sample consist of about 30 known
active G--M stars, with known photometric variations attributed to
rotational modulation of star spots, including both binaries and
--- supposedly --- single stars. Most of our sample stars belong
to the BY~Draconis variables, which are defined from their
variability due to changing star spot coverage \citep{bopp+1977}. Several
aspects need to be investigated to properly analyze the sample,
including (1) time-resolved RV monitoring and (2) line bisector
analysis, (3) monitoring of activity indexes, (4) determination of fundamental atmospheric parameters, and % finally 
(5) accurate photospheric abundance analysis. 
RV variations due to stellar acoustic oscillations, which for 
our purposes is an additional noise source, are effectively
filtered out by the long exposure times typically involved.

In this paper we will
investigate various tools available to perform reliable, accurate
and fast abundance analysis. Since this will have to be done in
detail for all stars in the sample, we are developing ways to
perform this task in as automated a way as possible, without employing
fully automated procedures yet \citep[e.g. minimum distance methods, see][]{allende2004},
although this is eventually the goal. Such automated tools will be of great utility in the future,
not least in view of upcoming facilities like STELLA \citep{stella2004}.

In this paper we present an analysis of the two
well-studied active stars \object{HD 27536} and \object{HD 216803},
deriving their fundamental atmospheric parameters and their
elemental abundances, as well as their activity indexes.
So far only
very few abundance analyses have been carried out for active
stars, mostly due to the complications involved \citep[see
e.g.][for recent applications to RS CVn
stars]{katz+2003,morel+2003}. 
We furthermore present a preliminary analysis of the time-resolved RV, bisector and
activity index data for \hdsyv .

 A future
paper will deal with the full sample of stars addressing the  RV, activity,
and line bisector variations and the statistical properties of the
sample, presenting a discussion of the effects of binarity on the
stellar activity.

\section{Observations}

\subsection{The target stars}

\hdsyv\ (EK Eri, HR1362) presents a special case. As shown by
\citet{strassmeier+1999}  it is a slowly rotating G8 giant with an
over-active chromosphere possibly due to a fossil Ap type magnetic
field, viewed at an inclination close to 90$^\circ$. The star has
a measured rotation period from long-term photometry of
306.9$\pm$0.4 days. Its Li abundance is close to solar. Yet its
magnetic activity level is very high and comparable to some RS~CVn
binaries.

\hdeen\ (TW PsA, HR8721, Gl879) is classified as a BY~Dra star
\citep{bopp+1977}. It is a K5 dwarf with a rotation
period of 10.3 days but with a small projected rotational velocity
of 2.6~\kms\ \citep{vogt+1983},
indicating an
inclination of approximately 30\degr . The star has been studied
recently by \citet{santos+2001}, who found atmospheric parameters
and metallicity and placed it in a sample of comparison stars
without planets.

Both stars are thought to be single.

\begin{figure}
\centering
\includegraphics[width=8.8cm]{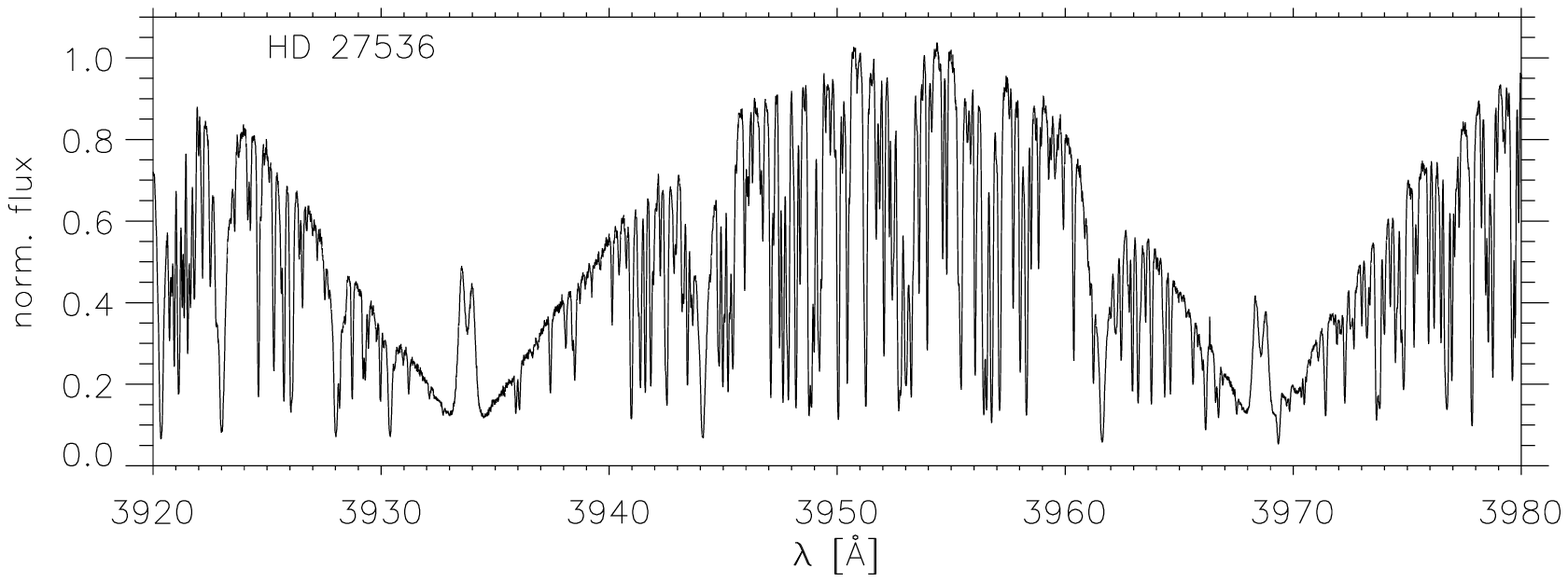}
\includegraphics[width=8.8cm]{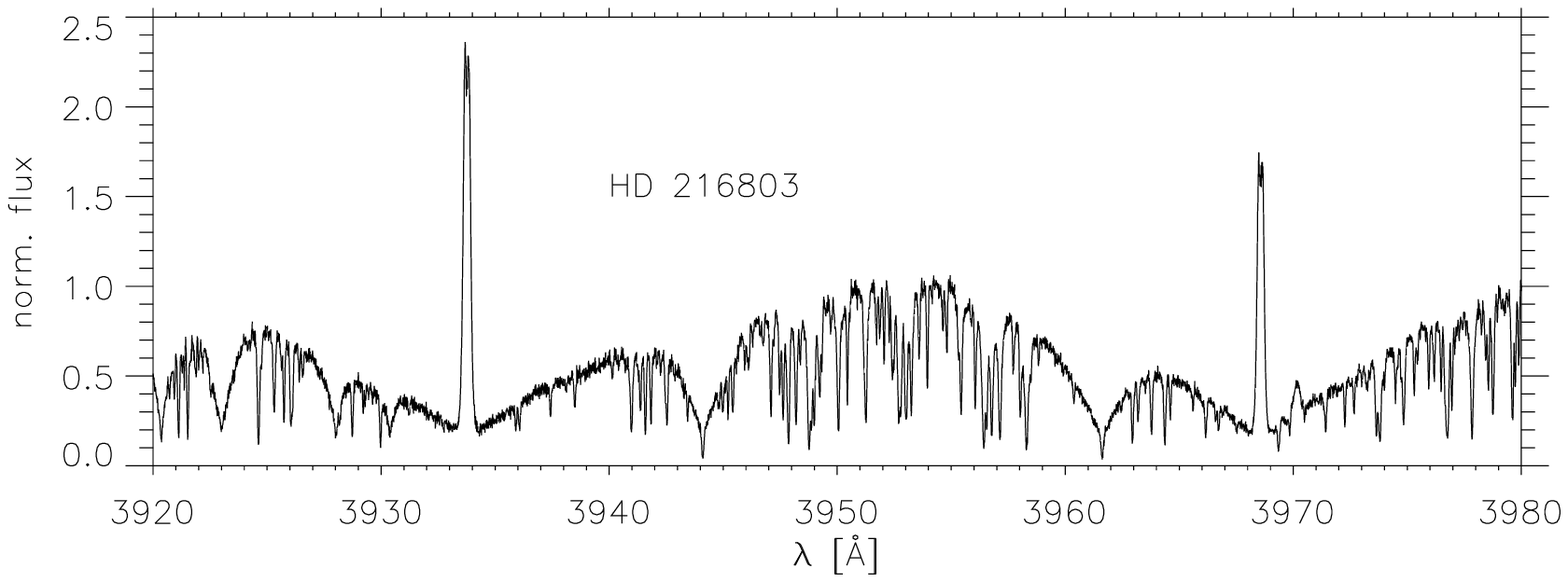}
\caption{\label{fig:caII}The \ion{Ca}{ii} H\&K lines region of
\hdsyv\ (top) and \hdeen\ (bottom). \hdsyv\ is a G8IV-III giant
while \hdeen\ is a K5V dwarf. }
\end{figure}

\subsection{New HARPS spectra}
\begin{table}
\caption{Log of the observations of \hdsyv\ and \hdeen . Exposure time is 400~s in all cases}\label{tab:obs}
\begin{tabular}{llr}\hline
Star & Date & S/N @ 650nm \\ \hline
\hdsyv\ &  2004-10-01 & 270 \\
        &  2004-10-02 & 420 \\
        &  2004-10-31 & 400 \\
        &  2004-11-23 & 245 \\
        &  2005-02-12 & 430 \\
        &  2005-03-17 & 230 \\
\hdeen\ &  2004-11-23 & 230 \\ 
\hline
\end{tabular}
\end{table}
Table~\ref{tab:obs} show the log of the observations on \hdsyv\ and \hdeen .
All spectra were taken with the
HARPS spectrograph at the ESO 3.6m telescope at La Silla
Observatory.
Exposure times were 400~s in all cases, which ensures that stellar acoustic noise is 
effectively averaged out.
In addition a high S/N solar spectrum was taken on the daylight
sky. All spectra were acquired in the simultaneous ThAr mode,
where a ThAr-lamp spectrum is recorded on the CCD alongside the
object spectrum. The reduction process of HARPS spectra is fully
automatic and extremely accurate due to the high intrinsic
stability of the spectrograph \citep{harps2004}. The final data
products is a wavelength calibrated 1D spectrum, with no
normalization or flux calibration performed, and a cross-correlation function (CCF),
computed for and averaged over all 72 spectral orders.  The formal error on
the wavelength calibration is $\sim 0.02$~m\,s$^{-1}$, while the
total RV uncertainty is $\sim1$~--~$2$~m\,s$^{-1}$ for all spectra.
 The spectral resolution is $R \sim 100,000$.
HARPS spectra cover the range 3780~--~6910~\AA , except for a gap between
5304~--~5337~\AA\ where order 115 falls between the two CCDs.

Both stars are clearly active, as evident from the \ion{Ca}{ii} H
\& K lines, shown in Fig.~\ref{fig:caII}.

\section{Methodology}\label{methods}

The two methods for abundance analysis described in the following
differ in their requirements onto the input spectrum. While method
1 (Sect.~\ref{meth1}) uses an automated procedure to fit the
continuum and the lines simultaneously, method 2
(Sect.~\ref{meth2}) uses an interactive procedure to normalize the
spectrum, followed by synthetic spectrum calculations.

\subsection{Abundance analysis: Method 1}\label{meth1}

The abundance analysis consists of the following steps:
measurement of the equivalent widths (EWs) over the full spectral
region, determination of a first guess at the basic parameters of
the star, calculation of the appropriate model atmosphere  and,
finally, the abundance analysis.  The method is the same as
employed by \citet{dall+2005b}, adopted from the procedures of
\citet{morel+2003} and \citet{bruntt+2002,bruntt+2004}.

The first step is the measurement of the EWs, which is
accomplished  using DAOSPEC\footnote{DAOSPEC has been written by
P.~B.~Stetson for the Dominion Astrophysical Observatory of the
Herzberg Institute of Astrophysics, National Research Council,
Canada.} \citep[][in preparation]{stetson+pancino2004}, which uses
an iterative Gaussian fitting and subtraction procedure to fit the
lines and the effective continuum. The lines are identified using
a list of lines from the VALD database
\citep[][]{kupka+1999,piskunov+1995}, where all lines deeper than
1\% of the continuum are included.  Different line lists for
different spectral types can be retrieved directly from the
database. In Sect.~\ref{linelist1}, we discuss the choice of line list in
more detail. For \hdsyv\ we used lines in the region
5000~--~6800~\AA, while for \hdeen\ the region 5500~--~6800~\AA\
was used. For bluer wavelengths the continuum determination
becomes uncertain, while the S/N decreases rapidly beyond
6800~\AA\ because of the decreasing instrument efficiency.
 Next, an initial estimate of $T_{\mathrm{eff}}$ is found using the line depth
ratios calibrated by \citet{kovtyukh+2003}. We approximate the
line depths by using the EWs directly, assuming the same width for all lines.  This is computationally
much easier, since DAOSPEC does not provide us with the line
depths. A more accurate determination of $T_{\mathrm{eff}}$ will
be derived later in the process. When used on the spectrum of the
Sun this procedure yields $T_{\mathrm{eff}}=5793 \pm 37$~K, hence
using the EWs instead of line depths prove accurate enough for a first estimate.  We next
adopt $\log g$ and microturbulence parameter $\xi_t$ for a
canonical ZAMS star, and assume solar metallicity as our starting
point. With this we then calculate the initial model, using the
ATLAS9 code adapted for Linux \citep{kurucz1993,sbordone+2004}.
With the measured EWs and the model, the abundances are calculated
using the WIDTH9 code \citep[][modified for PC by
V.~Tsymbal]{kurucz1993}, and compared line-by-line to Solar
abundances, calculated from a high S/N solar spectrum.  This last
step is crucial to avoid problems due to uncertain $gf$-values.
The parameters of the model used to calculate the solar abundances
are $T_{\mathrm{eff}} = 5778$~K, $\log g = 4.44$, and $\xi_t =
1.2$~km\,s$^{-1}$. 

Now the model parameters ($T_{\mathrm{eff}}$, $\log g$, $\xi$,
[Fe/H], and [$\alpha$/Fe]) are iteratively modified until
consistency is reached, defined by the following criteria:  (1)
That there are no trends of \ion{Fe}{i} abundance with EW,
wavelength or excitation potential, (2) that the abundances
derived from \ion{Fe}{i} and \ion{Fe}{ii} are the same, (3) that
the derived metallicity and $\alpha$-element abundances are
consistent with the input model.

The computational benefits of running these tools under Linux are
immense: On a \ion{Pentium}{iii} laptop the DAOSPEC computation of
800 EWs from the HARPS solar spectrum takes about three minutes,
while the calculation of a new ATLAS9 model takes two minutes. The
abundance calculation using the modified WIDTH9 on 400 iron lines
takes less than a minute.

\subsection{Abundance analysis: Method 2}\label{meth2}

The second method is based on the calculation of synthetic spectra
rather than just EW measurements. The software is called VWA which
has a graphical user interface (GUI). The method and the VWA software have been described in
detail by \citet{bruntt+2002,bruntt+2004} and the method compared to the
classical EW method by \citet{bikmaev+2002}. The abundance
analysis relies on atomic parameters from the VALD database
\citep[][]{kupka+1999,piskunov+1995} and uses modified ATLAS9
atmospheric models from interpolation in the grid published by
\citet{heiter+2002}. The least blended atomic lines are selected
automatically by VWA but additional lines can be chosen manually.
The abundance of each line is fitted iteratively by requiring that
the EW of the observed and computed spectrum agree. The wavelength
range used for matching of the EW is normally twice the FWHM of
the line, but if a line is partially blended, non-symmetrical, or
affected by a bad column in one of the wings, that part of the line
may be excluded when calculating the EW. When all lines have been
fitted the observed and synthetic spectra are compared. Offsets
due to wrong placement of the continuum are easily identified by
visual inspection in the GUI.

The VWA method requires that the spectra have been normalized. This is
done by manually defining continuum windows which we find by comparing
with a spectrum of the Sun \citep{hinkle+2000}, and
fitting a spline function through these points.

% -------------------- T1
\begin{table*}
\caption{Parameters for the two stars. $\log R_\mathrm{HK}$ and RV for \hdsyv\ are values 
from 2004-10-02, see Sect.~\ref{bisector} for details. See text for explanation of error estimates.
} \label{tab:parameters}
\centering
\begin{tabular}{lllllllllll}\hline
Star   &    $T_\mathrm{eff}$ [K] &  $\log g$   &  $\xi_t$        &
RV [km\,s$^{-1}$]  &  $v \sin i$ [km\,s$^{-1}$]  & [Fe/H] &
$\log R_\mathrm{HK}$ & $V-R$ \\ \hline
\hdsyv &  $5240\pm 35$ &  $3.55\pm 0.06$ & $1.2\pm 0.1$ &  $6.986\pm 0.002$ & $1.0 \pm 0.5$   & $+$0.09     & $-4.33 \pm 0.12$ & $0.67\pm0.03$ \\
\hdeen &  $4780\pm 50$ &  $4.70\pm 0.10$ & $1.1\pm 0.2$ &  $7.228\pm 0.002$ & $1.5 \pm 0.5$   & $-$0.05     & $-4.62 \pm 0.21$ & $0.95\pm0.06$ \\
\hline
\end{tabular}
\end{table*}

\subsection{Activity index}

We calculate the absolute emission fluxes in the  \ion{Ca}{ii}
H~\&~K lines, following the method by \citet{linsky+1979}: We
integrate the emission flux, $f_\mathrm{K}$, in the K line between
the K$_\mathrm{1V}$ and  K$_\mathrm{1R}$ points and we find
$f_\mathrm{H}$ in a similar way. This is then normalized by the
flux $f_{50}$ in the 50~\AA\ interval between 3925 and 3975~\AA,
and scaled to an absolute flux index:
\begin{equation}
\mathcal{F}_\mathrm{K} = 50 \frac{f_\mathrm{K}}{f_{50}} \mathcal{F}_{50},
\end{equation}
where $\mathcal{F}_{50}$ is defined from Eq.~3 of \citet{linsky+1979} for $V - R < 1.30$;
\begin{equation}
\log \mathcal{F}_{50} = 8.264 - 3.076 (V - R) .
\end{equation}
Following \citet{strassmeier+2000} we use $V - R$ colors computed
via  $B - V$ colors from the color-color relation of
\citet{gray1992}, since this minimizes the effects of the
rotational modulation of star spots \citep{strassmeier+1994}. In
order to calculate the activity index properly, we need to take
into account the contribution of the photosphere. Hence the
corrected flux indexes are (with a similar expression for the H
line)
\begin{equation}
\mathcal{F}^\prime_\mathrm{K} = \mathcal{F}_\mathrm{K} - \mathcal{F}^\mathrm{RE}_\mathrm{K} ,
\end{equation}
where $\mathcal{F}^\mathrm{RE}_\mathrm{K}$ is the index for a
radiative equilibrium atmosphere without a chromosphere, as given
in \citet{linsky+1979}. We then find the $R_\mathrm{HK}$ index as
the total chromospheric radiative loss in the H and K lines in
units of the bolometric luminosity:
\begin{equation}
R_\mathrm{HK} = \frac{\mathcal{F}^\prime_\mathrm{H} + \mathcal{F}^\prime_\mathrm{K}}{\sigma T_\mathrm{eff}^4} ,
\end{equation}
where we use our derived $T_\mathrm{eff}$.
Values of $B - V$ are found via SIMBAD and
checked for consistency with the derived $T_\mathrm{eff}$.  All
derived values are listed in Table~\ref{tab:parameters}.

\section{Results}

\subsection{Temperature, gravity, microturbulence}\label{teff}

\begin{figure}
\resizebox{\hsize}{!}{\rotatebox{-90}{\includegraphics{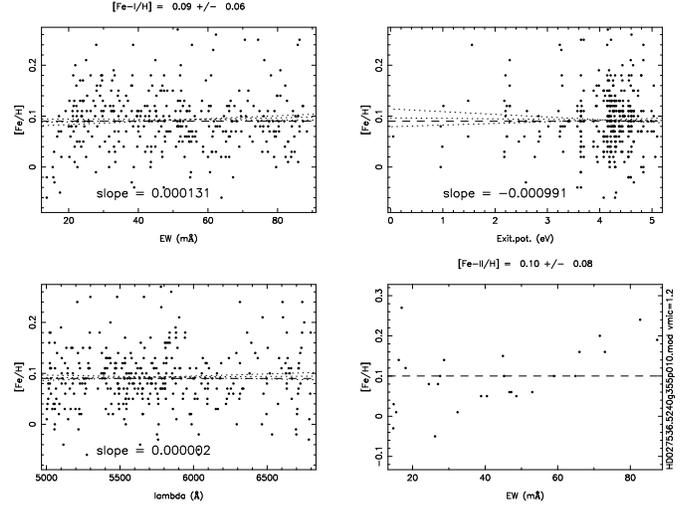}}}
\caption{\label{fig:diagnostics1}Method 1: The iron diagnostic plots used in
determining the atmospheric parameters of \hdsyv, using the
automatically selected Fe lines. Upper left plot and clockwise: Relative abundance of \ion{Fe}{i}
lines versus EW, same but versus excitation potential, relative abundance of \ion{Fe}{ii} lines
versus EW, \ion{Fe}{i} abundance versus wavelength of the line. The dashed line is the mean value at zero slope while
the dotted lines are the best linear fit and $\pm1\sigma$ fits.}
\end{figure}

Following the method outlined in Sect.~\ref{meth1} (Method 1) we
start with the calculation of EWs directly from the pipeline
produced 1D spectrum, which results in 1095 measured and
identified lines in \hdsyv.  Using the EW ratios instead of line
depth ratios, we find $T_{\mathrm{eff}}=5338$\,K\,$\pm60$\,K, and
adopt this value plus $\log g=4.40$ as our initial guess for the
model parameters.  After a few iterations we reach the model
parameters $T_{\mathrm{eff}} = 5240$~K, $\log g = 3.55$, $\xi_t =
1.2$~km\,s$^{-1}$, [Fe/H]~=~0.09, [$\alpha$/Fe]~=~0.0,  which
fulfills the convergence criteria (see Fig.~\ref{fig:diagnostics1}).  For \hdeen\ we arrive at
$T_{\mathrm{eff}}=4780$~K, $\log g =4.70$, $\xi_t
=1.1$~km\,s$^{-1}$, [Fe/H]~=~$-$0.05, [$\alpha$/Fe]~=~0.0.
The uncertainties are estimated from the difference between the best fit model and the model corresponding to the
$1\sigma$ fits in Fig.~\ref{fig:diagnostics1}.

\begin{table*}
\centering \caption{\label{tab:abund} The abundances ([M/H])
derived by our two methods for \hdsyv\ and \hdeen\ relative to the
measured abundances in a Solar spectrum.
The second column for
each star lists [M/Fe], i.e. the abundance within the star of each
element relative to iron.  The errors given are the RMS around the
mean.}
\begin{tabular}{llrrlrr|lrrlrr} \hline
 & \multicolumn{6}{c|}{\hdsyv} & \multicolumn{6}{c}{\hdeen} \\
 & \multicolumn{3}{c}{Method 1} & \multicolumn{3}{c|}{Method 2} & \multicolumn{3}{c}{Method 1} & \multicolumn{3}{c}{Method 2} \\
      & [M/H] & [M/Fe] & $N$ & [M/H] & [M/Fe] & $N$ & [M/H] & [M/Fe] & $N$ & [M/H] & [M/Fe] & $N$ \\ \hline
%Li & $+$0.10 & $+$0.01 &    2  &   & & \\
Al           & $+$0.18    & $+$0.09 &  1  &            &          &      &  $-$0.04     & $+$0.01 & 1  &              &   &   \\
Ca           & $+$0.19(6) & $+$0.10 & 30  & $+$0.16(1) & $+$0.07  &   3  &  $-$0.14(7)  & $-$0.09 & 7  &  $+$0.09(4)  & $+$0.08  &  3 \\
Si           & $+$0.12(7) & $+$0.03 & 37  & $+$0.07(5) & $-$0.02  &  18  &  $-$0.09(5)  & $-$0.04 & 12 &  $-$0.03(11) & $-$0.04  & 14 \\
\ion{Sc}{i}  &            &        &     &              &         &      &             &          &    &  $-$0.07      &         &  1 \\
\ion{Sc}{ii} & $+$0.11(2) & $+$0.02 &  7  & $+$0.11    & $+$0.02  &   2  &  $-$0.01(9)  & $+$0.04 & 5  &  $+$0.03     & $+$0.02  &  2 \\
\ion{Ti}{i}  & $+$0.14(7) & $+$0.05 & 30  & $+$0.13(6) & $+$0.04  &  16  &  $+$0.09(11) & $+$0.14 & 21 &  $+$0.10(15) & $+$0.09  & 24 \\
\ion{Ti}{ii} &            &        &      & $+$0.07(11) &         &  4   &             &         &      &              &  &   \\
V            & $+$0.27(9) & $+$0.18 & 17  & $+$0.25(8) & $+$0.16  &  11  &  $+$0.32(17) & $+$0.37 & 15 &  $+$0.25(20) & $+$0.24  & 11 \\
Cr           & $+$0.13    & $+$0.04 & 3   & $+$0.14(8) & $+$0.05  &   3  &  $-$0.02(7)  & $+$0.03 & 5  &  $+$0.06(12) & $+$0.05  &  7 \\
\ion{Fe}{i}  & $+$0.09(6) &         & 377 & $+$0.09(6) &          &  73  &  $-$0.05(7)  &         & 144 & $+$0.01(8)  &   & 78 \\
\ion{Fe}{ii} & $+$0.10(8) &         &  25 & $+$0.12(11) &         &  15  &  $-$0.02(8)  &         &  5 & $+$0.01     &   &  2 \\
Co           & $+$0.08(9) & $-$0.01 & 9   & $+$0.00    & $-$0.09  &   2  &  $+$0.00(10) & $+$0.05 & 6  &  $+$0.06(9)  & $+$0.05  &  5 \\
Ni           & $+$0.07(8) & $-$0.02 & 122 & $+$0.06(5) & $-$0.03  &  21  &  $-$0.08(7)  & $-$0.03 & 49 &  $-$0.03(9)  & $-$0.04  & 23 \\
\hline
\end{tabular}
\end{table*}

\begin{figure}
\centering
\includegraphics[width=8.8cm]{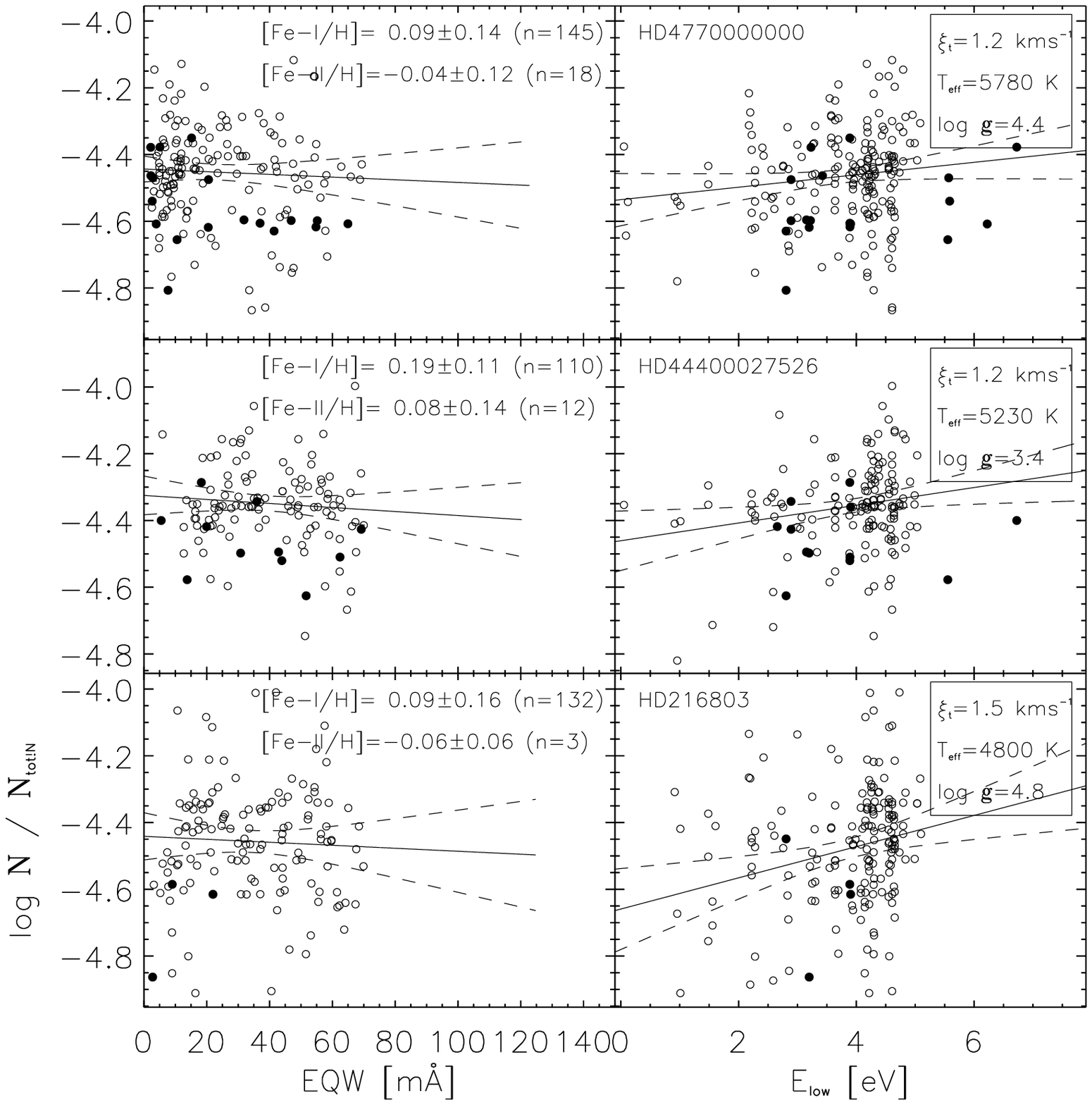}
\caption{\label{fig:sun}Method 2: Absolute Fe abundance found for the solar spectrum (top),
\hdsyv\ (middle) and \hdeen\ (bottom).
Open symbols are used for neutral iron lines and closed symbols are ionized iron lines.
In the left panels abundances are plotted versus EW while in the right panels
abundances are plotted versus excitation potential.
The parameters of the applied model is given in the insert boxes.}
\end{figure}

Following Method 2 outlined in Sect.~\ref{meth2},
we computed abundances for HD 27536, HD 216803 and
the Sun. For the spectrum of the Sun, we find iron abundances of
$\Delta A=+0.09$ and $-0.07$~dex for \ion{Fe}{i} and \ion{Fe}{ii}
respectively, where $\Delta A$ is the abundance relative to
standard value for the Sun, i.e. $\Delta A=\log N_{\rm Fe}/N_{\rm
tot} - (\log N_{\rm Fe}/N_{\rm tot})_\odot$. The value for the
solar abundance of Fe is $(\log N_{\rm Fe}/N_{\rm tot})_\odot$ and
is taken from \citet{grevesse+1998} who found $-4.54$. Our
measured abundances of \ion{Fe}{i} and \ion{Fe}{ii} are shown in
Fig.~\ref{fig:sun}. 
Open symbols are used for neutral lines and
filled symbols are for ion lines. The abundances are
plotted versus EW (left) and excitation potential (right). Note
that only lines with excitation potential in the range 2~--~5~eV
and EW in the
range 0~-~80~m\AA\ are plotted. 
In addition to
the large difference between abundances of \ion{Fe}{i} and
\ion{Fe}{ii} we see a significant correlation of \ion{Fe}{i}
abundance with excitation potential, an effect also seen when
using Method~1.  These discrepancies are explained
by a wrong temperature structure in the 1D-LTE ATLAS9 model for the Sun, as explained by \citet{shchukina+trujillo2001}.

\citet{allende+2004} also found discrepant results for the \ion{Fe}{i}/\ion{Fe}{ii} balance.
The discrepancy they found increased when going to late type stars and they
discussed whether spots in addition to the NLTE effects which are not included in
the 1D-LTE model atmospheres could explain this. In the present study we have
used the ion-balance to modify the $\log g$ value of the model atmosphere,
but this may not be the perfect approach. However, as we show later, a strictly differential
analysis, comparing the same lines between the stellar spectrum and the Solar spectrum, and
adopting the abundance determined from the most abundant ionic species, will largely
remove this problem.

An alternative approach to the determination of $T_\mathrm{eff}$ is the use of photometric color indices.
Using the calibrations of \citet{alonso+1999} for \hdsyv\ and the \citet{alonso+1996} calibrations for \hdeen\
with the Str\"omgren indices found by \citet{olsen1993} and $B-V$ colors from SIMBAD, we find
$T_\mathrm{eff}(b-y) = 5035$~K, $T_\mathrm{eff}(B-V) = 4998$~K for \hdsyv\ and
$T_\mathrm{eff}(b-y) = 4491$~K, $T_\mathrm{eff}(B-V) = 4438$~K for \hdeen , in both cases assuming
Solar metallicity. The uncertainties are around $120$ -- $140$~K for all values.
%
% hd27536:  b-y = 0.549, m1 = 0.312, c1 = 0.404, B-V = 0.90
% hd216803: b-y = 0.622, m1 = 0.630, c1 = 0.177, B-V = 1.10
Another temperature indicator is the Str\"omgren $\beta$ index.
However, no $\beta$ value is available for HD 27536. Instead, it was inferred from
the spectral type ($\beta = 2.54 \pm 0.01$) using the Str\"omgren indices
from \citet{olsen1993}. We then use the TempLogG software\footnote{e.g. http://ams.astro.univie.ac.at/templogg/}
\citep[see e.g.][]{kupka+bruntt2001}  which use the calibrations of
\citet{napiwotzki+1993}.
The result is $\log g = 3.1 \pm 0.7$ and T$_\mathrm{eff} = 5070 \pm 130$~K.
The interstellar reddening was found to be
$E(b-y) = 0.045 \pm 0.008$. 
The parameters of HD\,216803 fall
outside the valid range of the TempLogG calculations.
The discrepancy between color-derived and ion-balance derived T$_\mathrm{eff}$ points to problems
with the model atmospheres, the most obvious one being the lack of a chromosphere. As shown by \citet{morel+2003},
the use of ATLAS9 models without taking into account the chromosphere, 
results in an overestimation of T$_\mathrm{eff}$ and $\log g$. Their investigation
was based on the merging of an empirical chromosphere model and an ATLAS9 model, and although crude their results showed 
that the added chromospheric heating required a photospheric temperature several hundred K cooler. They also concluded that
the derived abundance ratios were not very sensitive to the addition of the chromosphere.  Lacking proper chromosphere models
we have employed the ATLAS9 models without modifications for consistency with past and future studies.

\begin{figure}
\centering
\includegraphics[width=8.8cm]{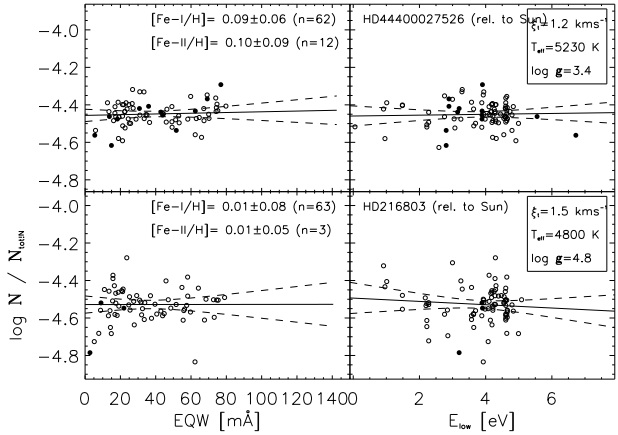}
%\vskip 0.2cm
\caption{\label{fig:fevwa}Method 2: Relative (to Solar) Fe abundance for \hdsyv\ (top row) and
\hdeen\ (bottom row) plotted versus EW (left column) and excitation
potential (right column). Same models and symbols as Fig.~\ref{fig:sun}.}
\end{figure}

To decrease the systematic errors from the NLTE effects and to reduce errors caused
by wrong $\log gf$ values we computed abundances in \hdsyv\ and
\hdeen\ relative to the abundances found for the same lines in the Sun.
This causes a decrease in the RMS of the abundance of \ion{Fe}{i} by
40\% (compare Figs.~\ref{fig:sun} and \ref{fig:fevwa}).
We only use lines with EW less than 90~m\AA\ and excitation potential
in the range 0 to 5~eV. The results of Fe for the best models of Methods 1 and 2 are shown in
Figs.~\ref{fig:diagnostics1} and~\ref{fig:fevwa} respectively.

\subsection{Chemical abundances}

\begin{figure}
\resizebox{\hsize}{!}{\includegraphics{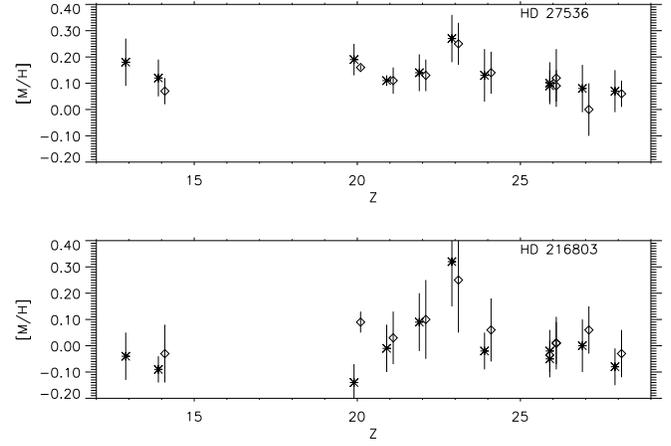}}
\caption{\label{fig:abund27+21}Relative abundances of \hdsyv\ and \hdeen\ from
Method 1 ($*$) and Method 2 ($\diamond$).  Note the overall good agreement, the 
deviating vanadium abundance and the discrepancy for calcium in \hdeen .}
\end{figure}

Abundances derived with our two methods are listed in Table~\ref{tab:abund}, and 
showed in Fig.~\ref{fig:abund27+21}. 
The errors quoted in the table are RMS around the mean abundance where a
sufficient number of lines were used. Additional errors are introduced by
uncertainties in the atmospheric parameters. Given the
uncertainties on $T_{\mathrm{eff}}$, $\log g$ and $\xi_t$, we
estimate an additional error of 0.05~dex which is to be added
quadratically to the RMS errors. Additional error sources are due
to NLTE effects, which are known to affect a few elements, like Fe (discussed in Sect.~\ref{teff}),
V, Mn and to a lesser extent Ti \citep{bodaghee+2003}. 
For the remaining elements the effects are unknown or uncertain.
EW measurement errors are likely significant for elements with only a
few measured lines i.e., Al, Sc and Cr. Small systematic errors could
be introduced by wrong continuum placement, but since we measure
all abundances relative to the Sun, this contribution is negligible.
On the other hand, random errors from the continuum treatment will
surely affect weak lines (e.g. \ion{Ti}{ii} and some lines of \ion{Fe}{ii}).

\begin{figure*}%\vspace{3cm}
\begin{centering}
\includegraphics[width=8cm]{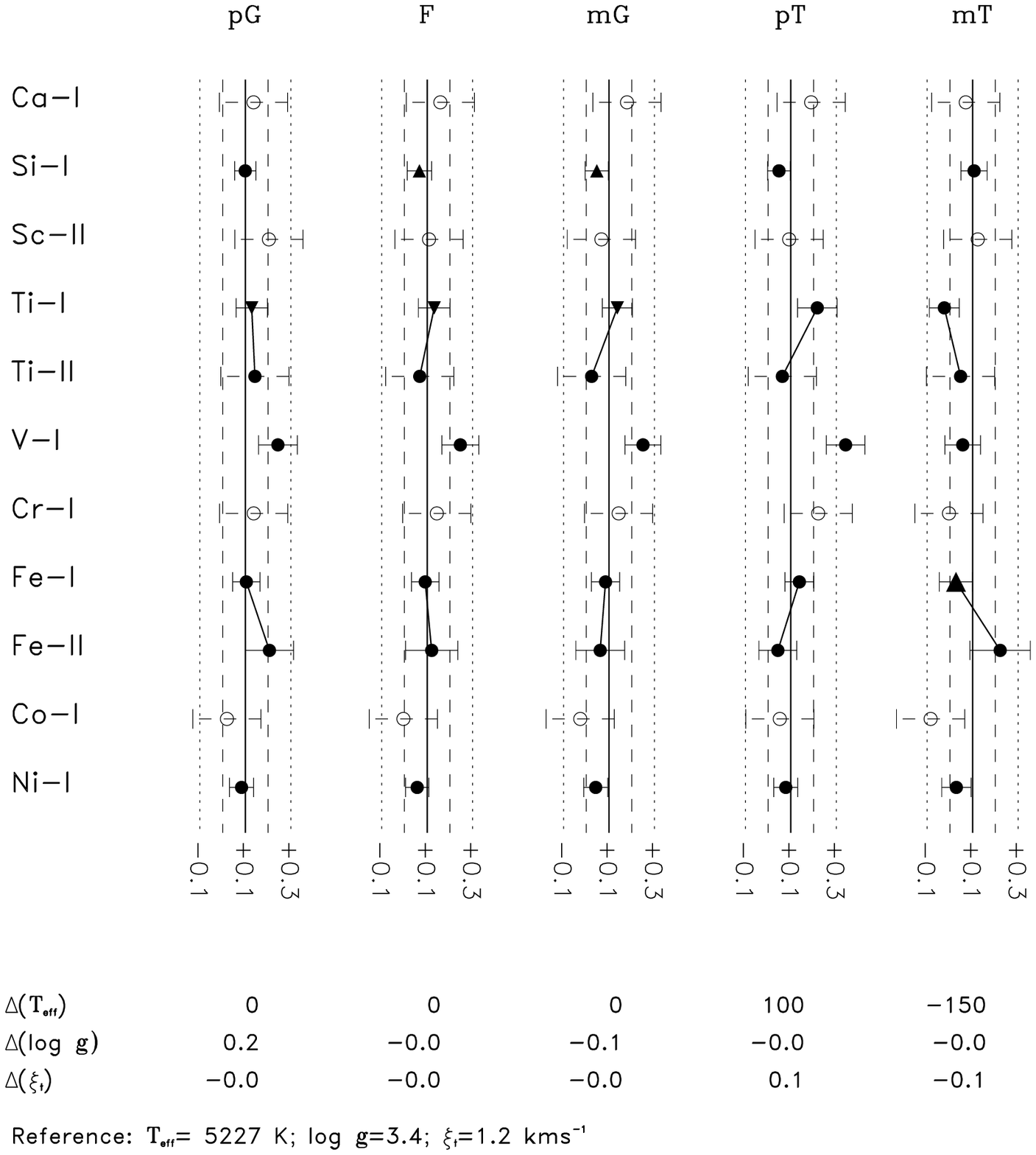}
\hskip 0.5cm
\includegraphics[width=8cm]{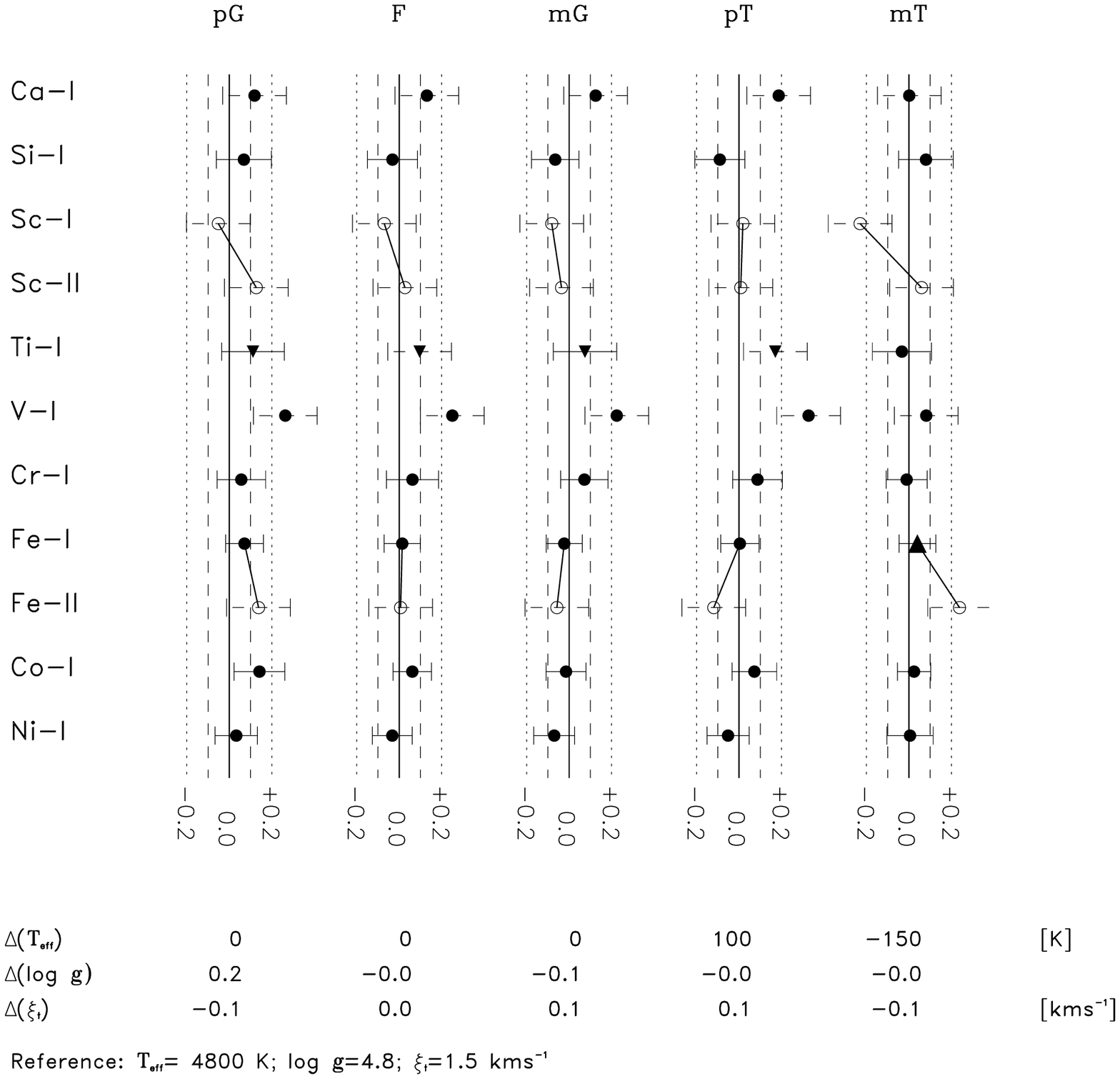}
\vskip 0.4cm
\end{centering}
\caption{Relative abundances found for different models of
HD~27536 (left panel) and HD~216803 by method 2. The preferred model is labeled
F while models with lower (higher) $\log g$ are labeled mG (pG)
and models with lower (higher) $T_{\rm eff}$ are labeled mT (pT). The differences with respect to model F
are listed below each model.
At the bottom of each panel the  $T_{\rm eff}$ and $\log g$ of model F are given (``Reference'').
Abundance values are relative to the Sun. \label{fig:vwa}}
\end{figure*}

The abundances of the main elements for several models are shown
in Fig.~\ref{fig:vwa}. The left panel is for HD\,27536 and the
right panel is for HD\,216803. The best model is labeled ``F''
while models with lower (higher) values of $\log g$ and
temperature are labeled ``mG/mT'' (``pG/pT''). The detailed
parameters of each model is given below each column. For example
the ``pT'' model of HD~216803 has $\Delta T=100$~K, $\Delta \log
g=0.0$ and $\Delta \xi_t=0.1$~\kms\ which is relative to the
reference model (i.e. model ``F'') at $T_{\rm eff}=4800$~K, $\log
g=4.8$ and $\xi_t=1.5$~\kms, thus the temperature of model ``pT''
is 100~K higher and the microturbulence is 0.1~\kms\ higher than
this. If there is a significant correlation (fitting a straight
line) of abundance and excitation potential, a triangle symbol is
used instead of a filled circle in Fig.~\ref{fig:vwa} (e.g.
\ion{Fe}{i} for the ``mT'' model). Open circle symbols are used if
less than five lines were used in the abundance determination
(e.g. for Sc in both stars).

A summary of the derived abundances of our preferred models ``F''
for HD 27536 and HD 216803 are given in Table~\ref{tab:abund}.

\subsubsection{Notes on individual abundances}

In the following, we discuss the determination of abundances of
individual elements and the error estimates.

\begin{figure}
\resizebox{\hsize}{!}{\includegraphics{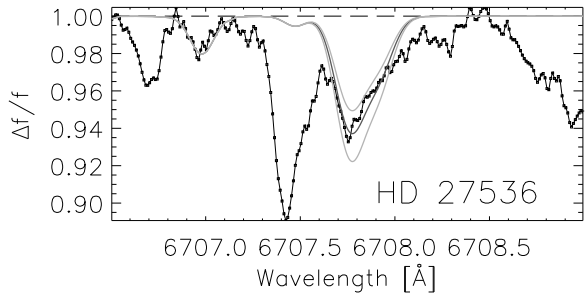}}
\caption{The lithium line region in \hdsyv\ with fits using three different abundances ($-$10.94, $-$10.84, $-$10.74).
Blending with \ion{Fe}{i} and \ion{V}{i}, but not with CN, has been taken into account.}
\label{fig:li-plot}
\end{figure}

\paragraph{The lithium abundance}
For measuring the Li resonance line at $\lambda\lambda$\,6707.8,
the automatic deblending performed by DAOSPEC proved inadequate,
and instead we used the IRAF\footnote{IRAF is distributed by the
National Optical Astronomy Observatories, which are operated by
the Association of Universities for Research in Astronomy, Inc.,
under cooperative agreement with the National Science Foundation,
U.S.A.} task \emph{splot} for the deblending as well as a spectral synthesis using VWA. The line is
actually a dublet split by fine structure. 
The major contribution
comes from the more abundant $^7$Li isotope, 
with blends from weak \ion{Fe}{i}, \ion{V}{i} lines, which is taken care of
with the VWA fitting, an example of which is showed in Fig.~\ref{fig:li-plot}.
The equivalent widths measured are
15.5~m\AA\ and 28.7~m\AA\ for HD\,27536 and HD\,216803, respectively.

The Li lines in the solar spectrum could not be reliably measured,
so we are 
using directly the atomic parameters for the lines. The
$gf$ values of the Li lines are among the most accurate for any
stellar lines \citep{smith+1998}, and can safely be used in an
absolute manner. The result is $A(\mathrm{Li}) = -10.84\pm 0.10$ and $-11.19\pm0.13$
for HD\,27536 and HD\,216803, respectively, the errors coming primarily from
uncertainty in the continuum placement and the influence from the weak \ion{Fe}{i}, \ion{V}{i}, and CN blends.
Identical results are obtained with splot and VWA.

The contribution from CN to the total EW has not been taken into account, since we cannot establish reliable
carbon and nitrogen abundances.  Only the two \ion{C}{i} $\lambda\lambda\,5052.167,5380.337$\,\AA\ lines can be 
measured, and they give deviating results. Assuming [C/Fe]~=~0 at most, we expect a contribution from
CN of up to a few m\AA, resulting in a possible overestimation of the lithium abundance by 0.1-0.2\,dex.

\paragraph{The $\alpha$-elements}
The abundances of the electron donor elements  Ca, Si and Ti merit
separate comments.  In active stars these elements are often
enhanced with respect to iron \citep[e.g.][]{morel+2003}, hence
the enhanced opacity due the H$^-$ absorption need to be accounted
for in the model calculations.

While [Si/Fe] and [Ti/Fe] generally agree between Methods 1 and 2,
there is some discrepancy concerning calcium for \hdeen, possibly due to
the abundance using Method 2 being based on only three lines.
 Taking an
unweighted mean of [Si/Fe], [Ti/Fe] and [Ca/Fe] for \hdsyv\ we get
[$\alpha$/Fe]~=~$+$0.06 ($+$0.03) from Method 1 (Method 2).
Excluding calcium, we find similarly [$\alpha$/Fe]~=~$+$0.05
($+$0.03) for \hdeen.  Within the uncertainties, this result is
compatible with no $\alpha$-enhancement for both stars, and we did
indeed use [$\alpha$/Fe]~=~0.0 in the model calculations.

\paragraph{Chromium}
Using Method 1 three \ion{Cr}{i} and three \ion{Cr}{ii}  lines
were used for \hdsyv. The \ion{Cr}{i} lines yielded $+$0.13 with a
spread of only 0.01~dex while the \ion{Cr}{ii} lines gave $+$0.05,
again with a spread of only 0.01~dex.   Assuming this difference to be real and
not just a statistical effect, we believe the difference to be due to
shortcomings of the model atmosphere, rather than reflecting
errors in $\log g$. 

\paragraph{Vanadium}
For both stars we find very high abundances of vanadium  ([V/Fe]) using
either method. Rather than a true overabundance, this may actually
reflect temperature inhomogeneities on the stellar surface,
indicating substantial spot coverage \citep[e.g.][]{gray+1991,strassmeier+sch2000}.
Vanadium lines are known to be strongly enhanced in sunspots \citep{wallace+1999}
since they have a very strong temperature
dependence. Alternatively, the cause may be NLTE effects, but we
do not find any trends of abundance versus EW, which would
indicate this.

\subsection{Line list comparisons}
\label{linelist1}
Several studies of stellar atmospheric abundances have employed
different line lists, all carefully crafted to suit the analysis
for which they are intended and this work is no exception. Our
line list is selected directly from the VALD database based on the
quality of the spectrum under investigation, i.e. the list will
depend on the S/N of the spectrum and the amount of blending, and
hence be a function of spectral type and rotational velocity.  We
have tested two different selection schemes: (1) From the full
line list we select manually the best non-blended lines from a
comparison with the spectrum under investigation, and (2) we use
the full VALD line list and subsequently reject outlying abundance
values.

\begin{figure}
\resizebox{\hsize}{!}{\rotatebox{-90}{\includegraphics{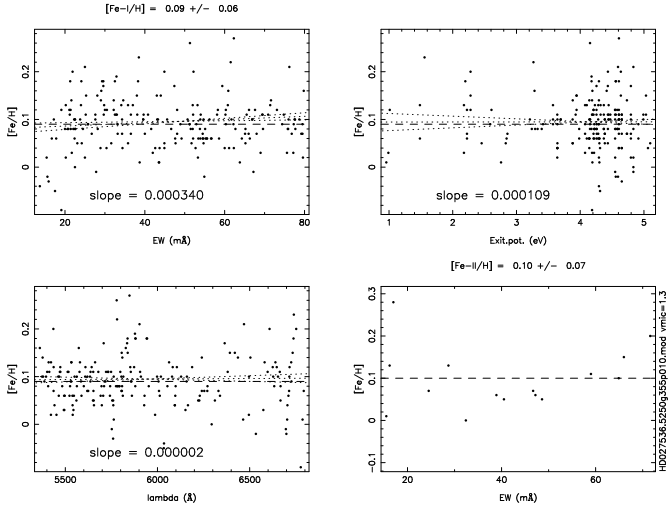}}}\\
\caption{\label{fig:diagnostics2}Same as
Fig.~\ref{fig:diagnostics1}, but using the manually selected line
list.}
\end{figure}

The two schemes yield very similar results when used to find the
stellar parameters for \hdsyv. In scheme 2 we use 346 \ion{Fe}{i}
and 31 \ion{Fe}{ii} lines after rejection iteration, while for
scheme 1 we selected 205 \ion{Fe}{i} and 15 \ion{Fe}{ii} lines.
The differences in atmospheric parameters for \hdsyv\ are $\Delta
T_\mathrm{eff} = 10$~K, $\Delta \log g = 0.00$, $\Delta \xi_t =
0.1$~km\,s$^{-1}$ between
%the parameters found using
the two line lists. In Figs.~\ref{fig:diagnostics1}
and~\ref{fig:diagnostics2} we show the diagnostic plots of the two
different line lists, both using Method 1 for the analysis.  The derived abundances differ by 0.02~dex at most, with
most elements differing less than 0.01~dex.  Hence we can conclude
that using all possible lines, relying on the statistical accuracy
of a sufficiently large number of lines works as well as a
manually selected set of lines.

\begin{figure}
\resizebox{\hsize}{!}{\includegraphics{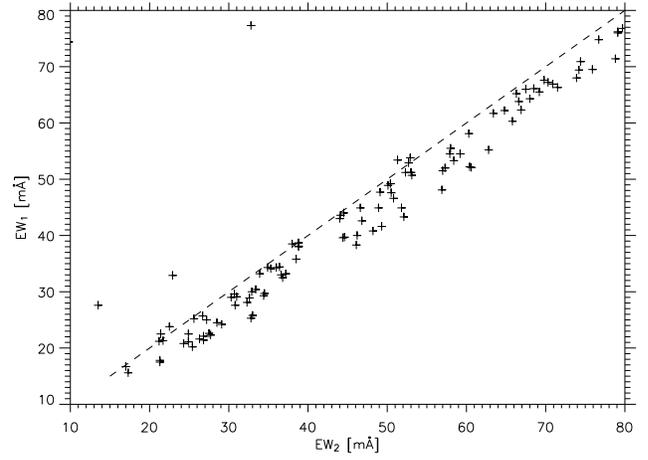}}\\
\caption{\label{fig:ews}The EWs as measured for 110 \ion{Fe}{i}
lines using DAOSPEC (EW$_1$) and VWA (EW$_2$). The dashed
line is the EW$_1$ = EW$_2$ relation. As can be seen, DAOSPEC EWs
are systematically lower than the synthesized.}
\end{figure}

While the two methods prove to be internally consistent,  we had some
concern about how they compare with each other. In
Fig.~\ref{fig:ews} we show the EWs measured using DAOSPEC (EW$_1$)
versus the EWs of the same lines using VWA (EW$_2$), and as can be
seen, there is a systematic offset between the EWs of the two methods.
%(\emph{ Maybe discuss the ``effective continuum'' approach by
%Stetson }).

\begin{figure}
\resizebox{\hsize}{!}{\rotatebox{-90}{\includegraphics{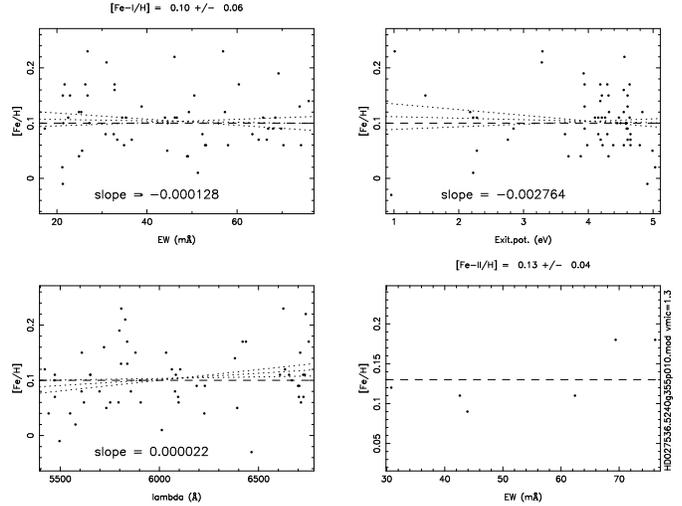}}}
\caption{\label{fig:5240bruntt} Iron diagnostic plot using the
EWs found by Method 2 with the model found by Method~1. See Fig.~\ref{fig:diagnostics1} for legend.}
\end{figure}

In order to investigate the effects this offset may have on the
model parameters, we calculated the abundances using the VWA EWs
with Method~1. In Fig.~\ref{fig:5240bruntt} we show the Fe
diagnostic plot for the VWA EWs with the model found by Method~1.

As can be seen, the agreement is excellent, and also the
abundances are all the same within 0.02~dex, which means that
either method of EW calculation leads to the same fundamental
model parameters and the same abundances --- provided they are
properly calibrated on the spectrum of the Sun. We refer the 
interested reader to the DAOSPEC manual\footnote{
Can be obtained from http://cadcwww.dao.nrc.ca/stetson/daospec/} for an excellent discussion of
continuum fitting and the problems involved.

   We necessarily
also calculated the absolute iron abundance of the Sun, again
using Method~2 EWs with Method~1; the results of which is
$A(\mathrm{Fe})_\odot = -4.47 \pm 0.13$, while using the DAOSPEC
EWs we find $A(\mathrm{Fe})_\odot = -4.53 \pm 0.14$, using exactly
the same set of \ion{Fe}{i} lines. The solar iron abundance from
\citet{grevesse+1998} is $A(\mathrm{Fe})_\odot = -4.54$, hence it
is not possible from this result to determine which EW
determination is the most correct (if indeed such a concept is meaningful), since both methods give
consistent results within the errors.

\subsection{Radial velocities, bisectors and activity indexes}\label{bisector}
The activity indexes for \hdsyv\ and \hdeen\ are listed in Table~\ref{tab:parameters}. The main error
source is the errors on $V-R$, while the error from measuring the fluxes in the spectra are around $0.02$ on $\log R_\mathrm{HK}$
for both stars.  \citet{strassmeier+1990} found $\log R_\mathrm{HK} = -3.85$ for \hdsyv . Taking into account the differences in
derived $T_\mathrm{eff}$ and $V-R$, the present result and the findings of \citet{strassmeier+1990} are similar: using their
parameters, our values of $\log R_\mathrm{HK}$ would shift by $+0.21$.

\begin{figure}
\resizebox{\hsize}{!}{\includegraphics{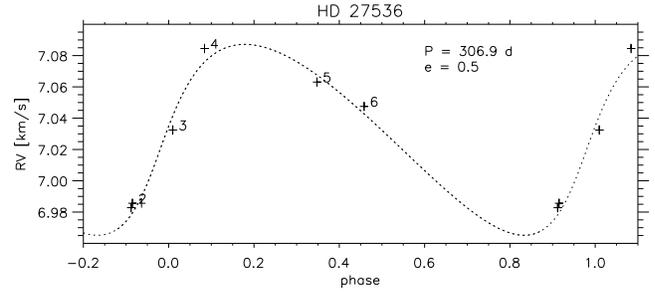}}\\
\caption{\label{fig:lc-27} RV variations measured in \hdsyv . The variation 
is consistent with the known 306.9~day period, but obviously long-term monitoring
is necessary. The dotted curve is a binary orbit fit with the known photometric period.}
\end{figure}

\begin{figure}
\resizebox{\hsize}{!}{\includegraphics{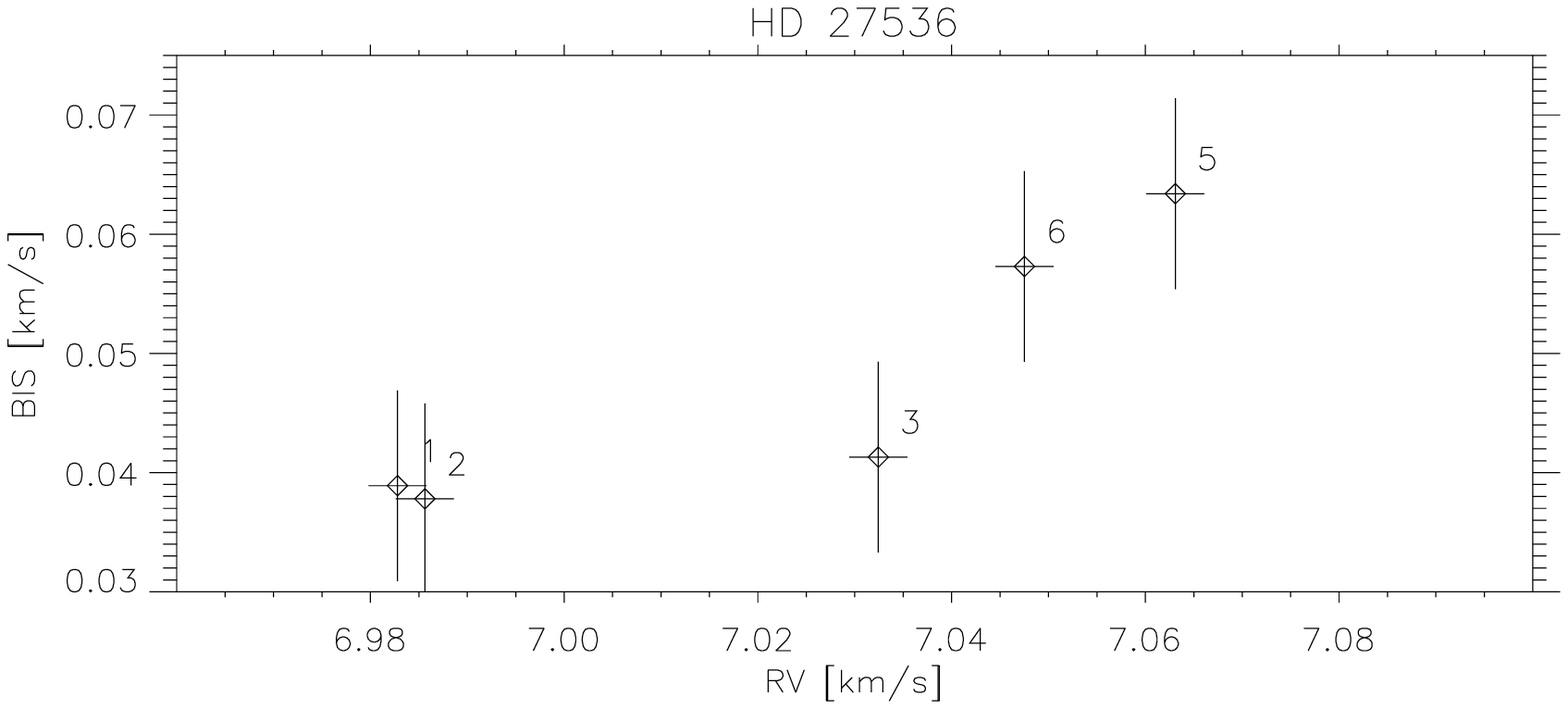}}\\
\resizebox{\hsize}{!}{\includegraphics{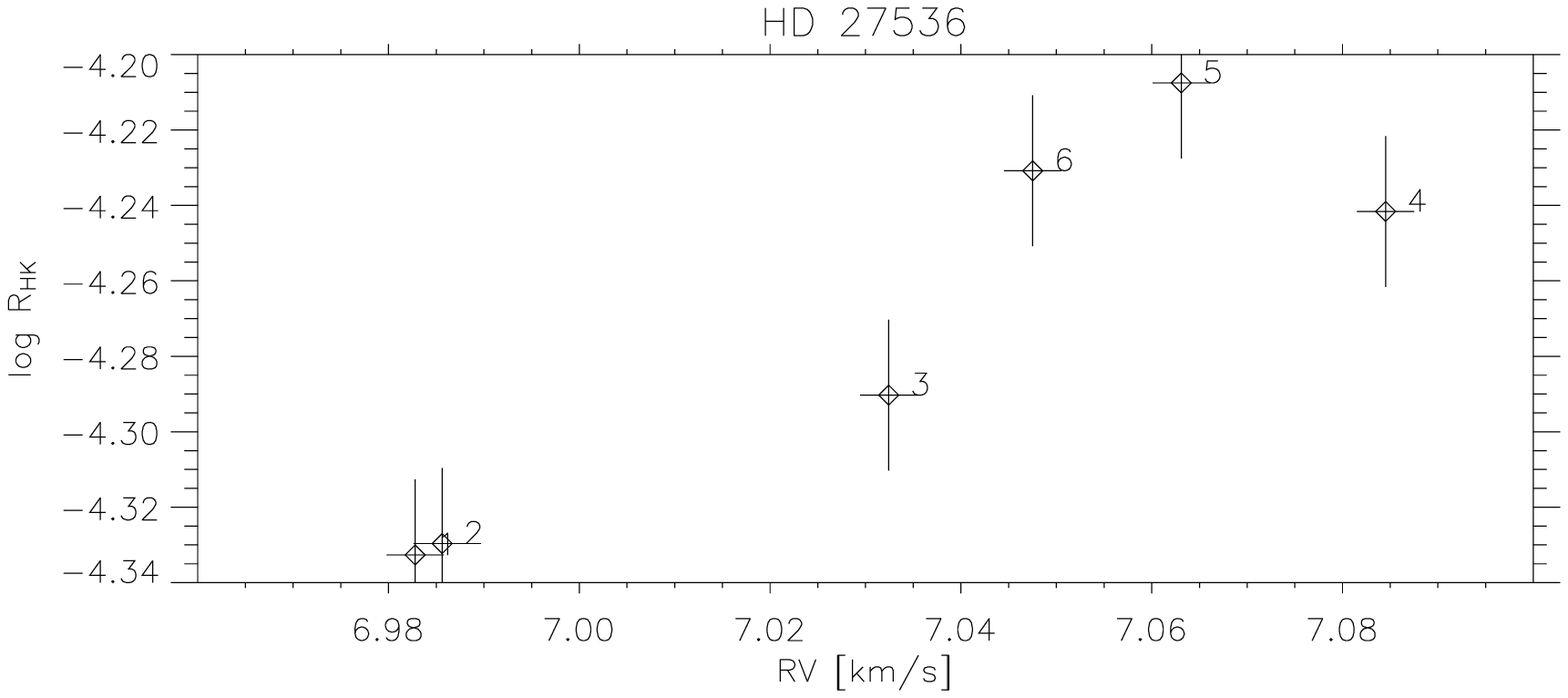}}\\
\resizebox{\hsize}{!}{\includegraphics{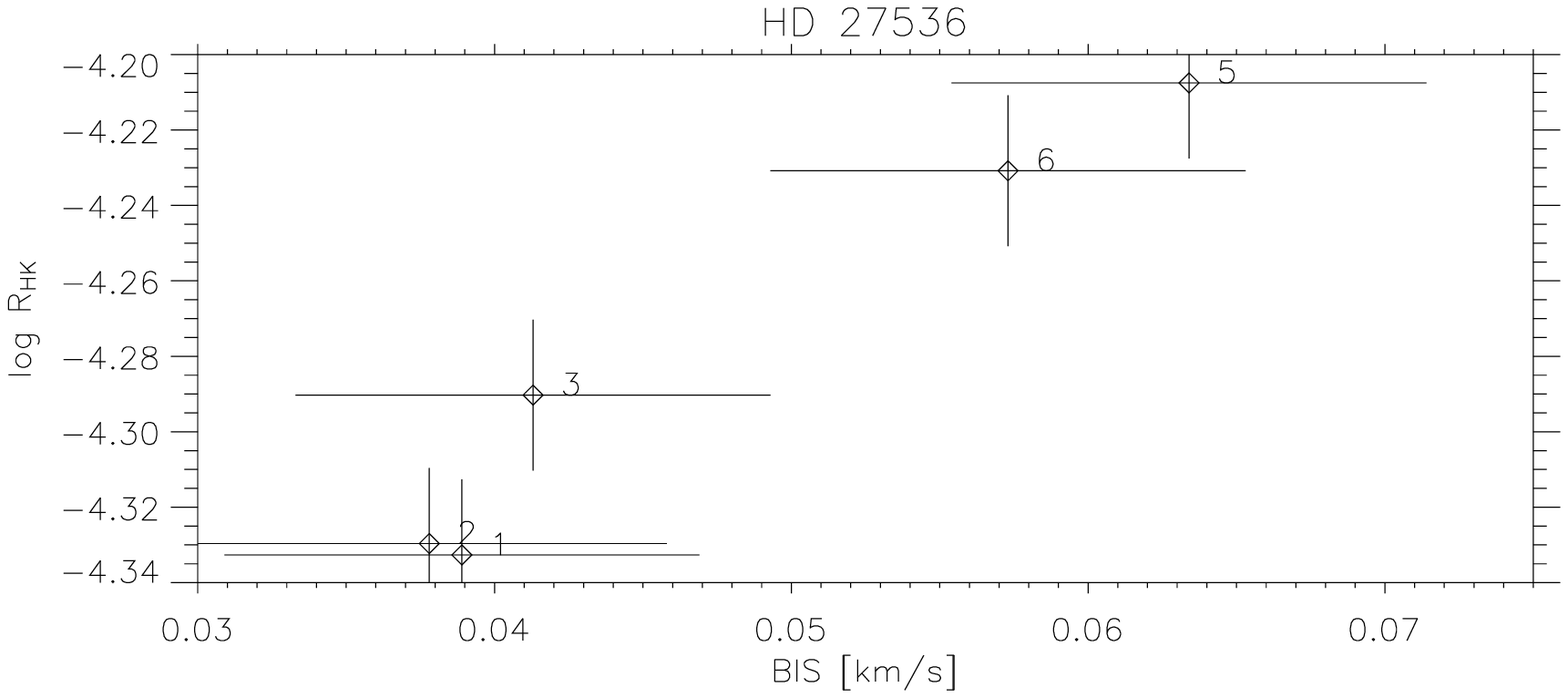}}\\
\caption{\label{fig:bis27} \emph{Upper panel:} Line bisector inverse slope (BIS; see text for definitions) vs RV.
\emph{Middle panel:} Activity index $\log R_\mathrm{HK}$ versus RV. 
\emph{Lower panel:} $\log R_\mathrm{HK}$ versus BIS. There seem to be a clear correlation between
activity level and RV, and between the BIS and the activity, pointing to changing activity as the cause of 
the RV variations. The error-bars on $\log R_\mathrm{HK}$ are only measurement errors i.e., not including uncertainties in $V-R$
and T$_\mathrm{eff}$, since these would affect all measurements equally.}
\end{figure}

For \hdsyv\ we have acquired six RV measurements with irregular intervals (see Table~\ref{tab:obs}). 
These are shown in Fig.~\ref{fig:lc-27}
along with an orbital fit using the known 306.9~d photometric period. While the measurements are consistent with the
period, it is clear that more measurements are needed.  To investigate whether this variation may be due
to a low-mass companion, we measured the line bisectors using the procedure of \citet{queloz+2001}. We measured
the bisector inverse slope (BIS) of the mean CCF of the spectra, and compared with the 
measured RV to check for correlations (Fig.~\ref{fig:bis27}, upper panel). As can be seen there seem to be a correlation
between the two quantities, indicating that the variation is internal to the star, probably due to activity
i.e. spots moving across the stellar disk.   Another indication of this comes from the activity index $R_\mathrm{HK}$,
which is shown in the middle panel of Fig.~\ref{fig:bis27} as a function of RV. This shows exactly the same behavior as the BIS,
strenghtening the interpretation that the RV variation is due to activity variations. The correlation between
the photospheric variation (BIS) and chromospheric ($R_\mathrm{HK}$) activity seem to be very strict
(Fig.~\ref{fig:bis27}, lower panel), in accordance with 
expectations \citep[e.g.][]{messina+2001}.

Obviously, part of the variation may be due to an unseen companion, hiding in the activity induced jitter.
Unfortunately, our measurements are too sparse to 
allow a definite conclusion.  It is worth noting that \citet{queloz+2001} found a negative correlation between
BIS and RV in the case of \object{HD 166435}, which allowed the authors to conclude that the RV variation was 
induced by activity. On the other hand, we find a positive correlation between BIS and RV, as did
\citet{santos+jvc2002} in the case of the visual binary \object{HD 41004}, which was ascribed to a brown dwarf
orbiting the fainter B component. Later, \citet{zucker+2004} identified the RV signature of a giant planet around
the A component.  Hence, in the case of \hdsyv\ we cannot exclude the possibility of companion induced
RV, activity and BIS variations, but more observations are evidently needed.  Even so, if we assume that the 
activity is indeed induced by an orbiting companion interacting magnetically with the star, we would expect
a RV ``jitter'' of 10--20~m\,s$^{-1}$ based on the calibrations of \citet{santos+2000}, i.e. consistent
with the observed variation, although their calibration strictly speaking only is valid for dwarfs.  
To disentangle any companion-induced pull from the activity jitter would thus require
several years of monitoring, since the period is so long and since the orbital pull and the induced activity
may have the same period.    
Assuming the period of the hypothetical companion to be equal to the photometric period of 306.9~d, and adopting
an inclination of $90^\circ$, we find a mass of $\sim 3.9$~M$_J$ orbiting in an
eccentric ($\epsilon \sim 0.5$) orbit with semi-major axis $\sim 1.2$~AU. This is under the assumption that all the RV variation is
due to an unseen companion. More likely, a companion would induce a smaller effect on top of the (larger)
activity induced jitter. 
Note also, that the photometric variation is sinusoidal, as shown by 
\citet{strassmeier+1999}, which is hard to reconcile with the eccentric ``orbit'' suggested by the RV data.


\section{Conclusions}

In this paper we have presented a pilot study of active stars, presenting and discussing the tools
to be used on a larger sample. 

We have presented results on the two photometrically variable, late-type active stars \hdsyv\ and \hdeen,
deriving their fundamental atmospheric parameters and photospheric elemental abundances from high-resolution
spectra.   We accomplished this task using two different methods, which produced results in good agreement
with each other, giving credibility to the derived parameters and abundances.  Furthermore, we have
showed that, despite the obvious differences in derived EW between the two methods, and despite the 
differences in line lists employed, we derive essentially the same results.   Both methods have a GUI, 
which greatly increases user-friendliness. Especially Method 1 has been developed with a view to
automation, and the ultimate goal is to be able to determine atmospheric parameters and abundances
for a great number of spectra without any human intervention.

While the agreement between the two methods is very tight for \hdsyv, larger differences resulted for
\hdeen\ ($\Delta$[M/H]$ \sim 0.06$). 
Even so, the abundances relative to iron, i.e. [M/Fe], show near perfect agreement between
the two methods, except for calcium, which may be due to poor statistics.  

Our study demonstrates the importance of determining relative abundances with respect to the Sun, which minimizes any
errors introduced by uncertain $gf$ values and uncertainties in continuum treatment.
On the other hand, the presence of surface spots may influence the determination of $\log g$ through the ion balance,
wherefore it may be better to use different approaches, or select lines based on their sensitivity to spots. 

Due to the high intrinsic stability of the HARPS spectrograph, a detailed analysis
of the RV and bisector variations was possible, using the limited data set of \hdsyv.
We find a RV variation, which is consistent with the known photometric period, although our coverage
is not long enough to judge whether this is indeed the true period, or if the variations are periodic at all.
We find a remarkable correlation between the bisector inverse slope and the activity index $\log R_\mathrm{HK}$,
which both vary in phase with the RV, revealing a very tight correlation between the photospheric and chromospheric activity.
The fact that we find a positive correlation may hint at the existence of an unseen companion, but our
phase coverage is not yet good enough to explore this possibility.

% acknowledgements
\begin{acknowledgements}
We are very  grateful to C.~Lovis for providing the line bisector analysis software, and to C.~Melo for calculating
the orbital elements. This research has made use of the SIMBAD database,
operated at CDS, Strasbourg, France.
\end{acknowledgements}

\bibliographystyle{bibtex/aa}
\bibliography{3539}

\end{document}